\documentclass[aps,pra,twocolumn,superscriptaddress,nofootinbib]{revtex4-2}
\usepackage{graphicx}
\usepackage{amssymb}
\usepackage{amsmath}
\usepackage{bm}
\usepackage{xcolor}
\usepackage{verbatim}
\usepackage{footnote}

\usepackage{hyperref}
\hypersetup{
     colorlinks=true,
     linkcolor=blue,
     filecolor=blue,
     citecolor=black,
     urlcolor=black,
     }
\usepackage[all]{hypcap}

\begin{document}

\title{Random spin textures in turbulent spinor Bose-Einstein condensates}

\author{Jong Heum Jung}
\affiliation{Department of Physics and Astronomy, Seoul National University, Seoul 08826, Korea}

\author{Junghoon Lee}
\affiliation{Department of Physics and Astronomy, Seoul National University, Seoul 08826, Korea}
\affiliation{Center for Correlated Electron Systems, Institute for Basic Science, Seoul 08826, Korea}

\author{Jongmin Kim}
\affiliation{Department of Physics and Astronomy, Seoul National University, Seoul 08826, Korea}

\author{Y. Shin}
\email{yishin@snu.ac.kr}
\affiliation{Department of Physics and Astronomy, Seoul National University, Seoul 08826, Korea}
\affiliation{Center for Correlated Electron Systems, Institute for Basic Science, Seoul 08826, Korea}
\affiliation{Institute of Applied Physics, Seoul National University, Seoul 08826, Korea}

\date{\today}

\begin{abstract}
We numerically investigate the stationary turbulent states of spin-1 Bose-Einstein condensates under continuous spin driving. We analyze the entanglement entropy and magnetization correlation function to demonstrate the isotropic nature of the intricate spin texture that is generated in the nonequilibrium steady state. We observe a $-7/3$ power-law behavior in the spin-dependent interaction energy spectrum. To gain further insight into the statistical properties of the spin texture, we introduce a spin state ensemble obtained through position projection, revealing its close resemblance to the Haar random ensemble for spin-1 systems. We also present the probability distribution of the spin vector magnitude in the turbulent condensate, which can be tested in experiments. Our numerical study highlights the characteristics of stationary turbulence in the spinor BEC system and confirms previous experimental findings by Hong {\it et al.} [Phys.~Rev. A {\bf 108}, 013318 (2023)].

\end{abstract}

\maketitle

\section{Introduction}

Quantum turbulence is a captivating phenomenon that arises in superfluids, characterized by chaotic flow with inviscidity and quantized circulation~\cite{vinen2002quantum,paoletti2011quantum,skrbek2021phenomenology}. The exploration of quantum turbulence has expanded to include atomic Bose-Einstein condensates (BECs), providing a unique platform to study this intriguing state~\cite{White2014,navon2016emergence,galka2022emergence}. In particular, BECs with internal spin degrees of freedom have enabled investigations of turbulence in spinor superfluids, which exhibit multiple velocity fields. The rich symmetries present in the order parameter manifold of spinor BECs give rise to the possibility of unconventional topological defects and different circulation rules~\cite{kawaguchi2012spinor,stamper2013spinor}, offering exciting prospects for the emergence of novel forms of turbulence~\cite{sadler2006spontaneous,kim2017critical,kang2017emergence,prufer2018observation,takeuchi2010binary,PhysRevA.85.033642,PhysRevA.88.063628,PhysRevA.84.023606,PhysRevA.94.023608}.

In a recent experiment by Hong {\it et al.}~\cite{hong2023spin} using spin-1 atomic BECs, it was observed that a stationary turbulent state emerged under the continuous application of an RF magnetic field. The oscillating magnetic field induces spin rotation and mixing, coupled with the dynamic instability of the system, leading to the formation of a nonequilibrium steady state with an irregular spin texture. Furthermore, in a specific driving condition, the magnitude of turbulence of the driven BEC was maximized in the spin sector to have an isotropic spin composition, indicating the presence of a fully developed spin-turbulent state. The investigation of the statistical properties of such a turbulent state presents an intriguing opportunity to explore complex far-from-equilibrium quantum phenomena.

In this paper, we present a numerical investigation focused on the spatial structure of the spin texture in the stationary turbulent state of the driven BEC system. We describe a spin-driving scheme to generate stationary turbulence and demonstrate its consistency with the experimental findings. Utilizing this driving scheme, we analyze the intricate spin texture in the nonequilibrium steady state after a prolonged driving time. We observe a $-7/3$ power-law behavior in the spectrum of the spin-dependent interaction energy~\cite{PhysRevA.85.033642, PhysRevA.88.063628} and find that the spin state ensemble obtained by
position projection is comparable to the Haar random ensemble for spin-1 systems. Our study elucidates previous experimental observations and highlights the isotropic nature and randomness of the spin texture in the stationary turbulent state.

\section{Model}

\subsection{Mean-field description}

We consider a homogeneous two-dimensional spin-1 BEC under an oscillating magnetic field. The magnetic field is given by 
\begin{equation}
    \mathbf{B}=\big(B_0 +\delta B_z(t)\big) \hat{\bm{z}} + B_\text{RF} \hat{\bm{x}}' \cos \omega t,
\end{equation}
where the first term is a uniform bias field with temporal fluctuations $\delta B_z(t)$ and the second term describes the RF oscillating field. The oscillating frequency $\omega$ of the RF field is equal to the Larmor frequency of $\omega_0=g_F \mu_\text{B} B_0/\hbar$, where $g_F$ is the Land\'{e} $g$-factor of the particle, $\mu_\text{B}$ is the Bohr magneton, and $\hbar$ is the Planck constant $h$ divided by $2\pi$. In a rotating frame, taking the rotating wave approximation, the mean-field Hamiltonian of the system is given by $H=H_0+H_{\textnormal{int}}$ with 
\begin{eqnarray}
    H_0 = \int d^2 \boldsymbol{r} \Big[ \Psi^{\dag}\big(-\frac{\hbar^2\nabla^2}{2m}-\hbar \delta(t) \text{f}_z + q\text{f}_z^2 - \hbar \Omega \text{f}_x \big)\Psi\Big], \nonumber \\
    H_{\textnormal{int}} = \int d^2 \boldsymbol{r} \Big[ \frac{g_n}{2}n^2+\frac{g_s}{2}\vert \boldsymbol{F} \vert^2 \Big], 
    \label{Hamiltonian}
\end{eqnarray}
where $\Psi = (\Psi_1, \Psi_0, \Psi_{-1})^\text{T}$ is the wave function of the spin-1 BEC, $m$ is the particle mass, $\textbf{f}=(\text{f}_x,\text{f}_y,\text{f}_z)$ denotes spin-1 matrices, $\delta(t)=g_F \mu_{\text{B}}\delta B_z/\hbar$, $q$ is the quadratic Zeeman energy, and $\Omega$ is the Rabi frequency of the RF driving. $H_\textnormal{int}$ represents the interaction energy of the system, where $g_n$ ($g_s$) denotes the particle (spin) interaction coupling constant and $n = \Psi^{\dag}\Psi$ ($\boldsymbol{F} = \Psi^{\dag} \textbf{f} \Psi$) is the particle (spin) density. Here, $\boldsymbol{F} = (F_x,F_y,F_z)$, with $F_{x,y,z}$ representing the magnetizations along the $x,y,$ and $z$ directions, respectively. The spin interaction is antiferromagnetic for $g_s > 0$ and ferromagnetic for $g_s<0$.

From the Hamiltonian, the dynamics of the BEC is described by the spin-1 Gross-Pitaevskii equation (GPE),
\begin{eqnarray}
    i \hbar \partial_{t} \Psi = \Big[ -\frac{\hbar^2}{2m}\nabla^{2}  -\hbar \delta \text{f}_z + q \text{f}_z^2 - \hbar \Omega \text{f}_x \nonumber \\
    + g_n n + g_s \boldsymbol{F} \cdot \textbf{f} -\mu \Big] \Psi,  
    \label{GPE}
\end{eqnarray}
where $\mu$ is the chemical potential of the condensate. 
The spin-1 BEC system has two characteristic length scales: density and spin healing lengths, $\xi_n = \hbar/\sqrt{2m\mu}$ and $\xi_s = \gamma \xi_n$ with $\gamma = \sqrt{g_n/|g_s|}$, respectively. The corresponding time scales are given by $t_n = \hbar/\mu$ and $t_s = \gamma^2 t_n$, respectively.

\subsection{Numerical simulation}

Based on the GPE, we numerically investigate the dynamics of the BEC in the experimental situation studied in \cite{hong2023spin}, where $\gamma=5.3$, $g_sn_0/h = 45$ Hz and $q/h= 47$ Hz. We model the field fluctuations $\delta B_z(t)$ as $\delta(t) = \delta_0 \sin{(\omega_\delta t+\phi)}$ with $\omega_\delta/2\pi = 60$~Hz, considering a typical experimental environment. In the experiment, the magnitude of field fluctuations was estimated to be approximately 1~mG and in our numerical study we set $\delta_0/2\pi = 1$~ kHz. $\phi \in [0,2\pi)$ is a random variable for different simulations. We set $\Omega/2\pi = 200$~Hz, which was found to generate a fully turbulent state in the experiment~\cite{hong2023spin}. The initial state is the easy-axis polar (EAP) state along the $z$-axis, that is, $\Psi_0 = \sqrt{n_0} (0,1,0)^\text{T}$ with $n_0=10^4/\xi_s^2$. Quantum fluctuations are taken into account using the truncated Wigner approximation~\cite{POLKOVNIKOV20101790}, for which each Bogoliubov excitation mode is populated by a fractionally occupied virtual particle, corresponding to half, in the EAP state with $q \to \infty$~\cite{blakie2008dynamics}. The size of the system is $l\times l$ with $l = 160\xi_s$, covered by a $1024\times1024$ grid of equally spaced points. 
The GPE is numerically solved using a relaxation pseudospectral scheme~\cite{besse2004relaxation, antoine2015gpelab}. The total particle number $N =\int d^2 \boldsymbol{r}~n(\boldsymbol{r})$ is preserved in the simulation.

\begin{figure*}[t]
    \includegraphics[width=17.5cm]{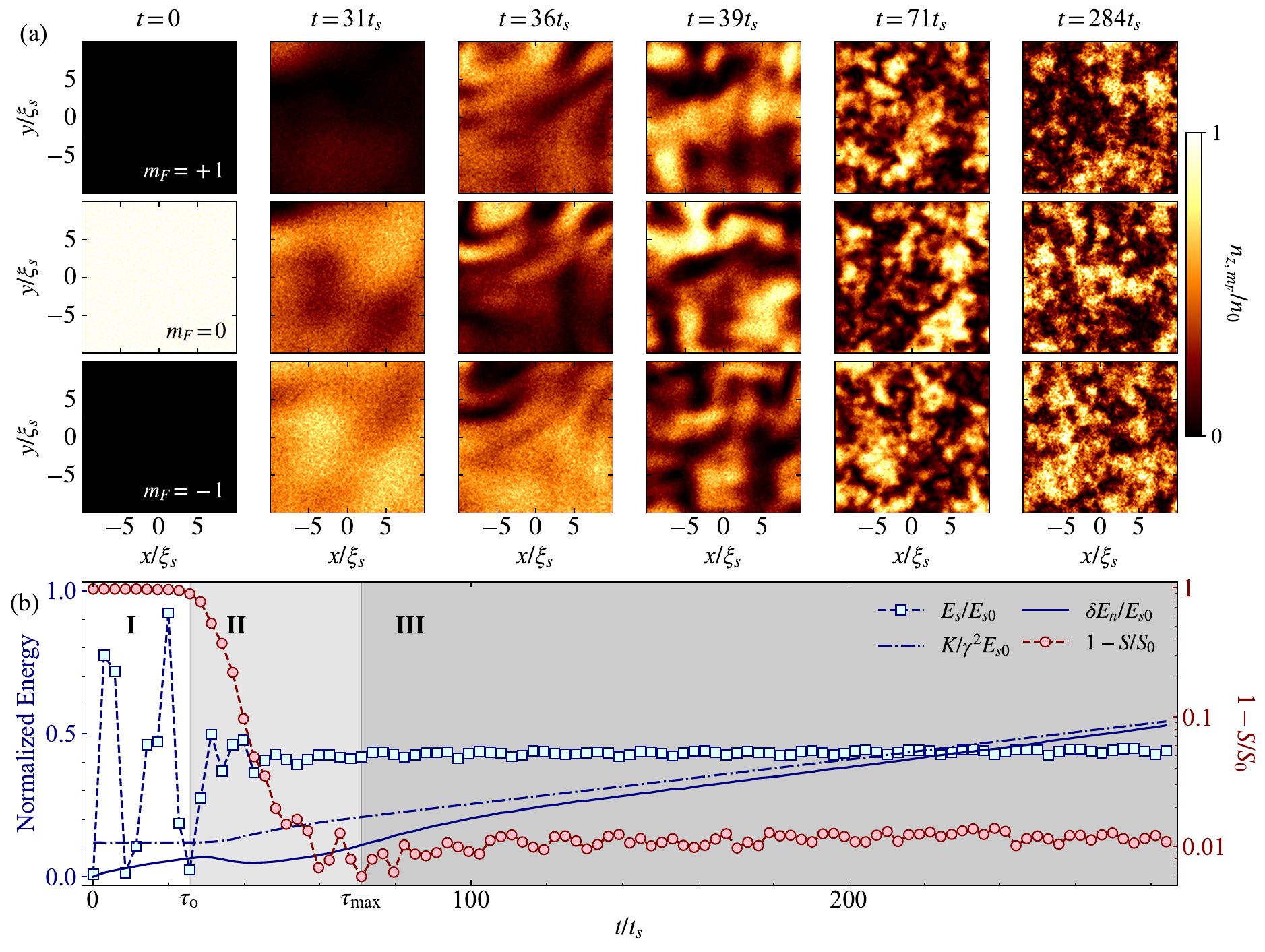}
    \centering
\caption{Emergence of stationary turbulence in a spinor Bose-Einstein condensate. (a-f) Density distributions of the $m_F$=$\pm1,0$ spin components, $n_{z,m_F}(\boldsymbol{r})$, at different times. (b) Time evolution of various energies and entanglement entropy of the driven BEC: spin interaction energy $E_s$ (blue square), kinetic energy $K$ (blue dashed-dotted line), increment of density interaction energy $\delta E_n$ (blue solid line), and spin entanglement entropy $S$ (red circle). Here $E_{s0}$ represents the characteristic spin interaction energy of the system, $\gamma^2=g_n/g_s$, and $S_0=\ln 3$ (see the text for details). The right vertical axis uses a logarithmic scale for $1-S/S_0$. Three distinct stages are identified: stage \textbf{I} ($t/t_s<\tau_\text{o}$) where $E_s$ oscillates with low $S$, stage \textbf{II} ($\tau_\text{o}<t/t_s<\tau_\text{max}$) where the entropy $S$ increases with developing irregular spin texture, and stage \textbf{III} ($t/t_s>\tau_\text{max}$) where the system reaches a stationary spin-turbulent state.
}
    \label{FIG1}
\end{figure*}

\section{Results and discussion}

\subsection{Emergence of stationary turbulence}

Figure~\hyperref[FIG1]{1(a)} displays the time evolution of the density distributions $n_{\alpha,m_F}(\boldsymbol{r})$ of the spin components.  Here, the index $\alpha \in\{x,y,z\}$ and $m_F\in\{+1,0,-1\}$ denote the quantization axis and the Zeeman sublevel, respectively. Initially, large-wavelength spin waves develop, indicating the system's dynamic instability, and the spin texture evolves to a more complicated spatial structure. Specifically, the spin texture breaks into smaller scales, revealing a characteristic energy cascade in turbulence, as observed experimentally~\cite{hong2023spin}. After a long time, the system maintains its complex spin texture. Note that such a complex spin texture does not occur without quantum noise in the initial state. It is quantum noise that triggers the dynamic instability~\cite{PhysRevA.76.043613,PhysRevA.75.013621} of the homogeneous system with spatially uniform parameters $q, \delta$, and $\Omega$.

In Fig.~\hyperref[FIG1]{1(b)}, we display the time evolution of various energies of the system, including the spin interaction energy $E_s$, increment of density interaction energy $\delta E_n$, and kinetic energy $K$, which are respectively calculated as,  
\begin{eqnarray}
    E_s =  \int  d^2 \boldsymbol{r} ~ \frac{g_s}{2} \vert \boldsymbol{F} (\boldsymbol{r}) \vert ^2, \nonumber \\
    \delta E_n = \int d^2 \boldsymbol{r} ~ \frac{g_n}{2} (n(\boldsymbol{r})^2-n_0^2), \nonumber \\
    K = \int d^2 \boldsymbol{r} ~ \Psi^{\dagger} (-\frac{\hbar^2}{2m}\nabla^2) \Psi.  
    \label{Energy}
\end{eqnarray}
The spin interaction energy initially oscillates a couple of times and then quickly converges to $E_s\approx 0.42E_{s0}$ for $t>50t_s$. Here, $E_{s0} = N g_s n_0/2$ %\textcolor{magenta}{XXX 1/2이 빠져서 추가하였습니다. XXX}
is the characteristic value of the spin interaction energy.
The damping of $E_s$ occurs when the spin texture becomes irregular and the steady value of $E_s$ indicates that the system enters a stationary turbulent state. 

We observe that both $\delta E_n$ and $K$ continue to increase gradually even after a long time. This means that while the spin texture reaches a dynamically steady state, the magnetic field driving incessantly injects energy into the system, and thus the energy transforms into kinetic energy and density fluctuations. 
%This is the thermalization process in TWA method~\cite{XXX}. 
In a real BEC system, this would result in heating of the driven system via energy dissipation and particle flux to a thermal component coexisting with the condensate. 
In fact, in the experiment carried out in \cite{hong2023spin}, it was observed that the thermal fraction of the system increases as spin turbulence is generated and reaches a new equilibrium value due to cooling by evaporation for the finite depth of the trapping potential. In our numerical simulation, the increase rate of the kinetic energy per particle for $t> 50t_s$ is almost constant and measured to be $\Gamma_{K}=7.9\times10^{-4}\mu/t_s \approx k_B \times 13$ nK/s, which is comparable to the estimated heating rate in the experiment. Here, $k_B$ is the Boltzmann constant. The damping rate of $E_s$ in its initial oscillation period is estimated to be $\Gamma_{E_s}\sim (E_{s0}/N)/50 t_s \approx 4\times10^{-4}\mu/t_s$, consistent with $\Gamma_{K}$.
Our numerical model is limited in its ability to study thermal relaxation adequately. In this study, we focus on characterizing the spatial structure of the intricate spin texture in the stationary turbulent state.

\subsection{Entanglement entropy}

We first investigate the spin-isotropic nature of the turbulence by calculating the entanglement entropy of the system. 
The BEC system can be described as a bipartite system consisting of spin and position. In other words, its Hilbert space can be represented as the tensor product of the spin space $\mathcal{H}_{\mathbb{S}}$ and the position space $\mathcal{H}_{\mathbb{R}^2}$.
According to Schmidt decomposition, the wave function of the BEC is expressed as
\begin{equation}
    \Psi(\boldsymbol{r})=\sum_{i=1,2,3} \alpha_i \phi_i (\boldsymbol{r}) \boldsymbol{\zeta}_i,
\end{equation}
where $\{\phi_i(\boldsymbol{r})\}$ and $\{\boldsymbol{\zeta}_i \}$ are the uniquely determined orthonormal base sets for the position and spin spaces, respectively, and $\alpha_i$ are referred to as Schmidt coefficients satisfying $\sum_{i=1}^{3} \alpha_i^2 = 1$. Then, the entanglement entropy of the system is given by $S=-\sum_{i=1}^{3} \alpha_i^2 \ln{\alpha_i^2}$. The entropy is maximized at $S_0=\ln{3}$ when $\alpha_i$ are all $1/\sqrt{3}$, and such a state with $S=S_0$ is called a \textit{maximally entangled state}.

We show the evolution of the entropy together with other energy quantities [Fig.~\hyperref[FIG1]{1(b)}]. As the spin interaction energy exhibits damping after a few initial oscillations, the entropy rapidly increases and reaches $S/S_0 \approx 0.99$ after $t \approx 60t_s$, indicating that the BEC system evolves into a maximally entangled state. In the experiment conducted in \cite{hong2023spin}, Hong {\it et al.} demonstrated that the stationary turbulent state has balanced populations over the three spin states, particularly for any quantization direction. This isotropic spin composition is the key feature of the maximally entangled state, consistent with our numerical results.

According to the behavior of $S(t)$, we identify three stages, namely \textbf{I}, \textbf{II}, and \textbf{III}, in the emerging process of turbulence: in \textbf{ I}, $S(t) \approx 0$ with coherent spin oscillations; in \textbf{II}, $S(t)$ increases with an irregular spin texture emerging; and in \textbf{III}, $S(t)$ reaches a steady value close to $S_0$, where the spin turbulence is fully developed. Transition times $\tau_\text{o}$ and $\tau_\text{max}$ are determined as $S(\tau_\text{o})=0.1 S_0$ and the time when $S$ is maximized, respectively. In our numerical results, $\tau_{\text{o}}\approx 25 t_s$ and $\tau_{\text{max}} \approx 71 t_s$. Note that the entropy is slightly reduced after $\tau_\text{max}$.

\subsection{Isotropic spin texture}

\begin{figure}[t]
    \includegraphics[width=8.6cm]{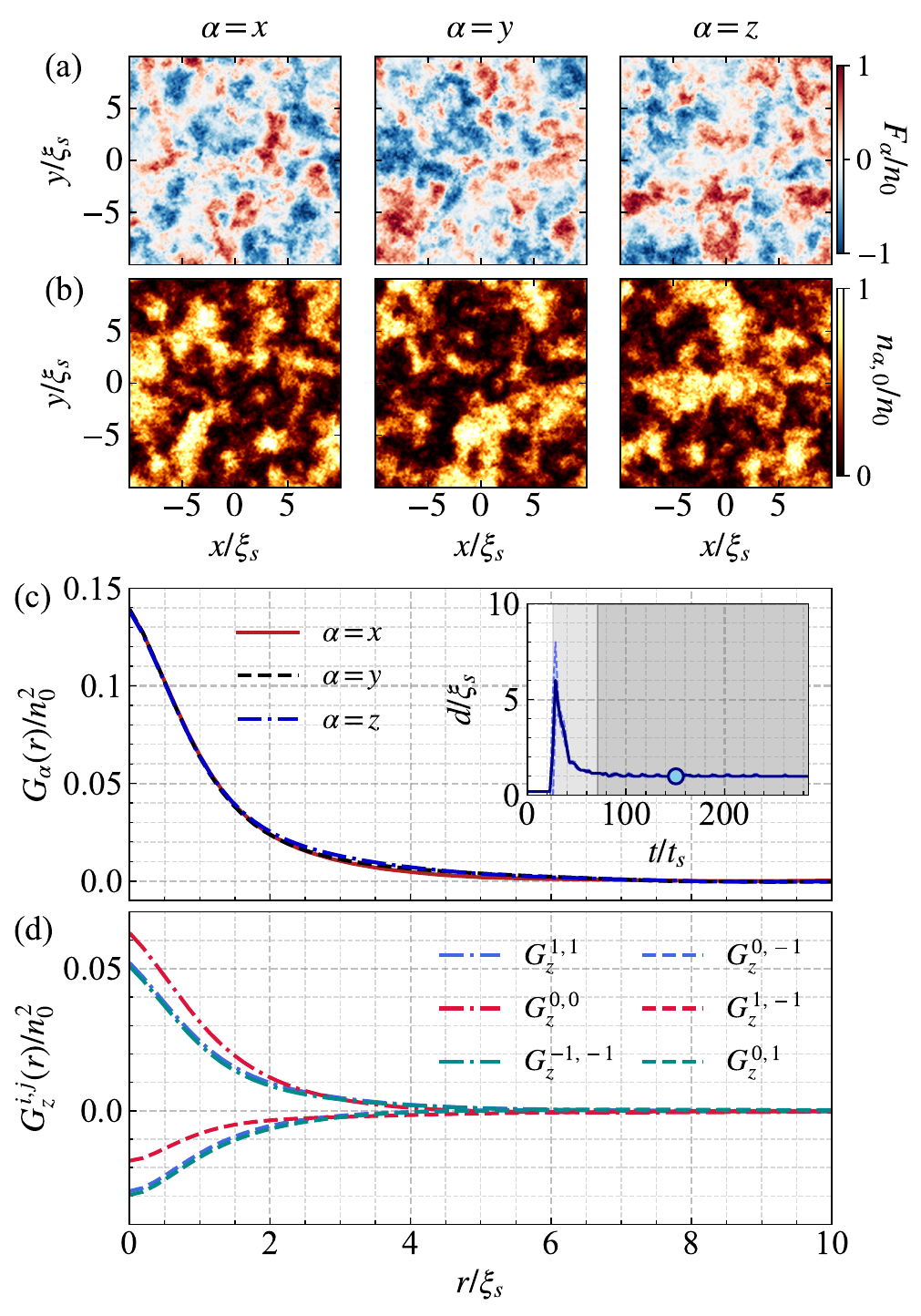}
    \centering
\caption{Isotropic spin texture of the turbulent spinor BEC. (a) Magnetization and (b) $m_F=0$ density distributions for $\alpha=x,y,z$ directions at $t = 150t_s$. (c) Magnetization correlation functions $G_{\alpha}(r)$ and (d) density-density correlation functions $G_{z}^{i,j}(r)$ for $i,j\in\{+1,0,-1\}$ at $t=150t_s$. The inset of (c) shows the time evolution of the averaged domain size $d$, which is determined from $G_{\alpha}(d_i)=G_{\alpha}(0)/2$. The blue-shaded region indicates the standard deviation of domain sizes $d_{\alpha}$ ($\alpha=x,y,z$). The white, light gray, and dark gray-shaded regions indicate the stage \textbf{I}, \textbf{II}, and \textbf{III} as in Fig.~\hyperref[FIG1]{1(b)}, respectively. The blue circle indicate the time point of $t=150t_s$.}
    \label{FIG2}
\end{figure}

In Fig.~\hyperref[FIG2]{2(a)}, we present the magnetization distributions, $F_\alpha(\boldsymbol{r})$, for $\alpha=x,y,z$ directions at $t=150 t_s$ when the spinor BEC is in a stationary turbulent state. To characterize the spatial structure of the irregular spin texture, we analyze the magnetization correlation function $G_{\alpha}(\boldsymbol{r})$, defined as  
\begin{eqnarray}
    G_{\alpha}(\boldsymbol{r}) = \langle F_{\alpha} (\boldsymbol{r}+\boldsymbol{r'})F_{\alpha}(\boldsymbol{r'}) \rangle_{\boldsymbol{r'}}-\langle F_{\alpha}(\boldsymbol{r'})\rangle^2_{\boldsymbol{r'}},
    \label{Correlation}
\end{eqnarray}
where $\langle \cdot \rangle_{\boldsymbol{r'}}$ denotes the averaged value of $\boldsymbol{r'}$. By angular averaging for $|\boldsymbol{r}|=r$, we obtain a one-dimensional function, $G_{\alpha}(r)$, shown in Fig.~\hyperref[FIG2]{2(c)}. These correlation functions decay as $r$ increases, and, remarkably, they exhibit an identical profile for all spin axes ($\alpha=x,y,z$), highlighting the isotropic character of the spin texture of the turbulent BEC. $G_{\alpha}(0)$ represents the variance of the magnetization along the $\alpha$ axis, and for $\langle F_\alpha\rangle=0$, it is related to the spin interaction energy as $G_\alpha(0)=\frac{1}{3} (E_s/E_{s0}) n_0^2$. 

We determine the domain size $d_{\alpha}$ as the radius at half maximum, that is, $G_{\alpha}(d_{\alpha}) = \frac{1}{2}G_{\alpha}(0)$. Their time evolution is shown in the inset of Fig.~\hyperref[FIG2]{2(c)}. In stage \textbf{III} (dark grey), the domain sizes become constant at $d_\alpha \approx \xi_s$, providing further evidence that the driven BEC reaches a steady state in its spatial structure. 

We further analyze the density-density correlation function $G_{\alpha}^{i,j}(r)$ ($i,j=+1,0,-1$), which is obtained by the angular averaging of 
\begin{eqnarray}
    G_{\alpha}^{i,j}(\boldsymbol{r})  = \langle n_{\alpha,i} (\boldsymbol{r}+\boldsymbol{r'})n_{\alpha,j}(\boldsymbol{r'}) \rangle_{\boldsymbol{r'}} \nonumber \\ 
    -\langle n_{\alpha,i}(\boldsymbol{r'})\rangle_{\boldsymbol{r'}} \langle n_{\alpha,j}(\boldsymbol{r'})\rangle_{\boldsymbol{r'}}. 
    \label{Correlation2}
\end{eqnarray}
As observed in $G_\alpha(r)$, the functions $G_{\alpha}^{i,j}(r)$ exhibit isotropic behavior, showing identical profiles for all the spin axes. The results are shown in Fig.~\hyperref[FIG2]{2(d)}, specifically for the $\alpha=z$ axis.

For the special case in which the three components are equivalent in terms of their interactions, $G_{\alpha}^{i,i}=G_{\alpha}^{\text{intra}}$ and $G_{\alpha}^{i,j}=G_{\alpha}^{\text{inter}}$ for $i\neq j$. Additionally, if the total density is spatially uniform, i.e., $\sum_j n_{\alpha,j}=n_0$, then $\sum_{j} G_{\alpha}^{i,j}=0$, yielding $G_{\alpha}^{\text{inter}}=-G_{\alpha}^{\text{intra}}/2$. Using these relations and considering $F_{\alpha}=n_{\alpha,+1}-n_{\alpha,-1}$, it is suggested that $G_{\alpha}=G_{\alpha}^{1,1}+G_{\alpha}^{-1,-1}-2G_{\alpha}^{1,-1}=3 G_{\alpha}^{\text{intra}}$. The numerical results shown in Figs.~\hyperref[FIG2]{2(c)} and \hyperref[FIG2]{2(d)} are qualitatively explained by this random three-component model. However, $G_{\alpha}^{0,0}$ is slightly higher than $G_{\alpha}^{1,1}$ and $G_{\alpha}^{-1,-1}$. We attribute this to the antiferromagnetic interactions of the system, which induce phase separation between the $m_F=0$ and $m_F=\pm 1$ components, while the $m_F=\pm 1$ components remain miscible.

\subsection{Spin energy spectrum}
\begin{figure}[t]
    \includegraphics[width=7cm]{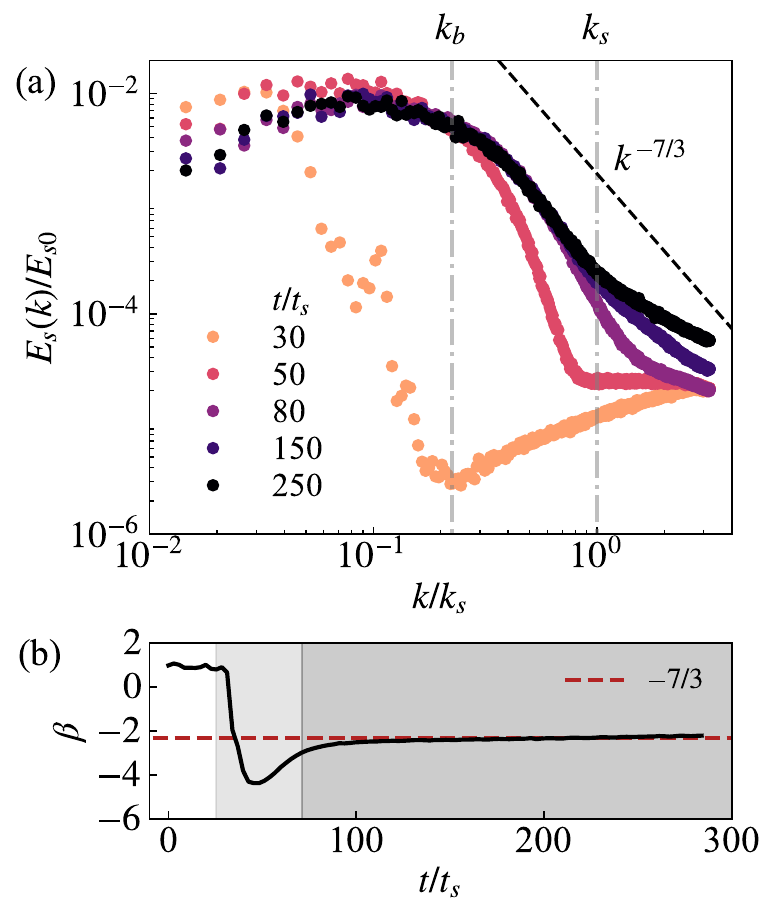}
    \centering
\caption{Spin energy spectrum $E_s(k)$ for various times $t$. The $x$ and $y$ axes have a logarithmic scale.  The black dashed line shows $k^{-7/3}$. The exponent $\beta$ was determined from a power-law fit to the region of $k_b< k< k_s$ and the inset shows $\beta$ as a function of $t$. Gray dashed-dotted lines indicate $k=k_b$ and $k=k_s$.}
    \label{FIG3}
\end{figure}

Turbulence is conventionally characterized by the energy spectrum of the velocity field. In the inertial range over which energy is transferred with negligible dissipation, a characteristic scaling behavior such as the Kolmogorov $-5/3$ power-law has often been observed in classical and quantum fluids. In Refs.~\cite{PhysRevA.85.033642, PhysRevA.88.063628}, Fujimoto {\it et al.} reported a set of numerical results showing that spin turbulence can be developed in the spin-1 condensate system and the spin interaction energy exhibits a steady $-7/3$ power-law spectrum in various driving situations. They also provided an analytical argument for its occurrence within the wave number range of $k_b<k<k_s$, where $k_s = 2\pi/\xi_s$ and $k_b = k_s/\sqrt{2}\pi$, assuming that the magnitude of the spin vector is small.

Motivated by these previous results, we investigate the spin energy spectrum of our BEC system under continuous magnetic driving. To analyze the spectrum of $E_s$, we transform the spin density vector $\boldsymbol{F}$ as
\begin{eqnarray}
    \boldsymbol{F}(\boldsymbol{r}) = \frac{1}{l} \sum_{\boldsymbol{k}} \tilde{\boldsymbol{F}}(\boldsymbol{k}) e^{i\boldsymbol{k}\cdot\boldsymbol{r}} 
    \label{spinvec}
\end{eqnarray}
and obtain the spectrum of spin interaction energy as 
\begin{eqnarray}
    E_s(k) = \frac{g_s}{2} \sum_{k<\vert \boldsymbol{k} \vert<k+\Delta k} \vert \tilde{\boldsymbol{F}}(\boldsymbol{k}) \vert^2,
    \label{Esk}
\end{eqnarray}
where $\Delta k = \frac{2\pi}{l}$ is a grid size in $k$-space~\cite{PhysRevA.85.033642}.
In Fig.~\hyperref[FIG3]{3(a)}, we present the energy spectra $E_s(k)$ at different evolution times. These results demonstrate that as energy is injected into the low-$k$ region, a propagation front emerges. It is characterized by a steep slope and progresses towards the high-$k$ region. Once the system reaches a stationary turbulence state for $t>\tau_\text{max}$, we observe saturation of the energy spectrum within the range of $k_b<k<k_s$. In particular, the spectrum exhibits a scaling behavior that closely approximates a power-law scaling with an exponent of $-7/3$.

We estimate the power-law exponent $\beta$ by fitting the spectrum in the range of $k_b<k<k_s$ to a power-law function of $E_s(k)=E'_{s} k ^{\beta}$. In Fig.~\hyperref[FIG3]{3(b)}, the variation of $\beta$ is shown as a function of $t$. At $t=0$, $\beta$ is equal to 1 owing to the contribution of quantum noise. In the early part of stage \textbf{II}, where $S$ undergoes a sudden increase [Fig.~\hyperref[FIG1]{1(b)}], $\beta$ experiences a rapid decrease to the lowest value of $\beta\approx-4$, indicating that the energy front passes through the $k$ window. Subsequently, $\beta$ increases and eventually saturates at $\beta \approx -7/3$, indicating the establishment of a stationary state. The inset also reveals a slow increase in the scaling exponent $\beta$ for larger $t$. This behavior arises from the continuous flow of energy from the low-$k$ to high-$k$ region without any energy dissipation channel, resulting in the accumulation of spin energy at high-$k$. This is also linked to the observed progressive increase at $E_K$ and $\delta E_n$ in Fig.~\hyperref[FIG1]{1(b)}.

\subsection{Spin state ensemble}

\begin{figure}[t]
    \includegraphics[width=8.6cm]{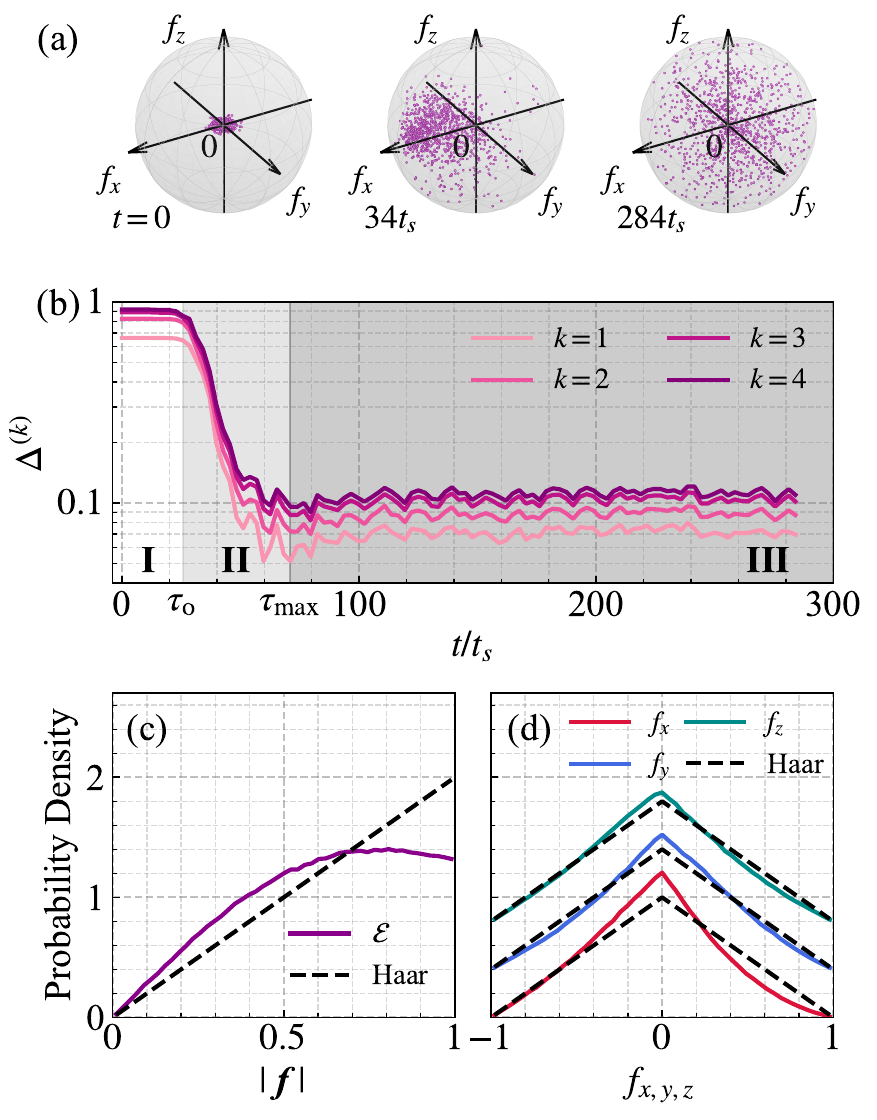}
    \centering
\caption{Spin statistics of the turbulent spinor BEC. (a) Scatter plots of the spin state ensemble $\mathcal{E}$ of the BEC in the coordinate space of $(f_x,f_y,f_z)$ at $t= 0$, $34t_s$, and $284t_s$. 1000 spin states were randomly chosen from the ensemble. The sphere indicates the surface of $\vert \boldsymbol{f}\vert=1$. (b) Trace distances $\Delta^{(k)}$ ($k=1,2,3,4$) between the $k$-th moments of $\mathcal{E}$ and the Haar random ensemble as functions of time. The $y$-axis has a logarithmic scale. (c) Probability density distributions of $\vert \boldsymbol{f} \vert$ and (d) $f_{x,y,z}$ at $t=284t_s$. The dashed lines indicate the corresponding distributions from the Haar random ensemble. In (d), the data for $f_y$ and $f_z$ were plotted with offsets of $0.4$ and $0.8$, respectively, for clarity.}
    \label{FIG4}
\end{figure}

To gain further insight into the spin randomness of the stationary turbulent state, we adopt the concept introduced in Refs.~\cite{PRXQuantum.4.010311,ippoliti2022dynamical,PhysRevLett.128.060601} and consider a projected ensemble of spin states. As the Hilbert space of the BEC system is the tensor product of the spin space $\mathcal{H}_{\mathbb{S}}$ and position space $\mathcal{H}_{\mathbb{R}^2}$, through projective position measurement of the BEC, we obtain an ensemble $\mathcal{E}$ of pure spin states supported on $\mathcal{H}_{\mathbb{S}}$,
\begin{eqnarray}
    \mathcal{E}= \Big\{ \Big ( \frac{n(\boldsymbol{r})}{N}, \boldsymbol{\zeta}(\boldsymbol{r}) \Big) \Big\}. 
    \label{projectedEnsemble}
\end{eqnarray}
Here, each spin state in the ensemble is associated with a local position $\boldsymbol{r}$, i.e., $\boldsymbol{\zeta}(\boldsymbol{r})=\frac{1}{\sqrt{n(\boldsymbol{r})}} \Psi(\boldsymbol{r})$, and weighted by $\frac{n(\boldsymbol{r})}{N}$, which is the probability of finding the spin state in the BEC at position $\boldsymbol{r}$. 
This projected ensemble carries more information than the conventional reduced density matrix~\cite{choi2023preparing}, allowing a more comprehensive characterization of the statistical properties of the system. For instance, the $k$-th momentum of the ensemble is defined as 
\begin{eqnarray}
    \rho_{\mathcal{E}}^{(k)} &=& \mathbb{E}_{\boldsymbol{\zeta} \sim \mathcal{E}} [ (\boldsymbol{\zeta} \boldsymbol{\zeta}^{\dagger} )^{\otimes k} ] \nonumber \\
    &=& \sum_{\boldsymbol{r}\in\mathbb{R}^2} \frac{n(\boldsymbol{r})}{N} \big( \boldsymbol{\zeta}(\boldsymbol{r}) \boldsymbol{\zeta}^{\dagger}(\boldsymbol{r}) \big)^{\otimes k},
    \label{moment}
\end{eqnarray}
where $\mathbb{E}_{\boldsymbol{\zeta} \sim \mathcal{E}}$ denotes ensemble-averaging over the elements $\boldsymbol{\zeta}$ in $\mathcal{E}$~\cite{PRXQuantum.4.010311}, and then the $k$-th moment of the arbitrary observable $O$ for the ensemble is expressed in terms of $\rho_{\mathcal{E}}^{(k)}$ as follows: 
\begin{equation}
    O^{(k)} = \mathbb{E}_{\boldsymbol{\zeta} \sim \mathcal{E}} [ (\boldsymbol{\zeta}^{\dagger} O \boldsymbol{\zeta} )^k]= \text{tr}(\rho^{(k)}_{\mathcal{E}}O^{\otimes k}).
    \label{observable}
\end{equation} 
The first moment of the ensemble corresponds to the reduced density matrix, providing information about the expectation values of any observable, while the second moment of the ensemble contains information about the variance of the observable. 

Following the method described in~\cite{PRXQuantum.4.010311}, we estimate the randomness of the spin state ensemble from its comparison to the Haar random ensemble $\mathcal{E}_\text{Haar}$, which is a unitarily-invariant ensemble such that the statistics of the spin-1 system has a maximally entropic distribution at the level of the Hilbert space $\mathcal{H}_{\mathbb{S}}$~\cite{ippoliti2022dynamical}. 
The distinguishability between both ensembles is measured with the trace distance $\Delta^{(k)} $ in their $k$-th moments, \begin{eqnarray}
    \Delta^{(k)} \equiv \frac{1}{2}\vert\vert \rho_{\mathcal{E}}^{(k)}-\rho_{\text{Haar}}^{(k)} \vert\vert_{1},
    \label{QuantumStateDesign}
\end{eqnarray}
where $\vert\vert \cdot \vert\vert_{1}$ denotes the trace norm. We call the ensemble {\it quantum state $k$-design} if $\Delta^{(k)}=0$~\cite{ambainis2007quantum}.\footnote{Because of the property of trace distance that $\Delta^{(j)} \leq \Delta^{(k)}$ for any $j \leq k$~\cite{ippoliti2022dynamical}, the quantum state $k$-design is also a quantum state $j$-design for $j \leq k$.} 
In Fig.~\hyperref[FIG4]{4(b)}, we plot the trace distances $\Delta^{(k)}$ ($k=1,2,3,4$) as functions of time. The temporal behavior of $\Delta^{(k)}$ is similar to that of $1-S/S_0$ in Fig.~\hyperref[FIG1]{1(b)}, clearly demonstrating the spin randomization as turbulence is generated in the system. In stage \textbf{III}, $\Delta^{(4)}\leq0.12$ and the system is close to the quantum state 4-design.

\begin{figure}[b]
    \includegraphics[width=8.6cm]{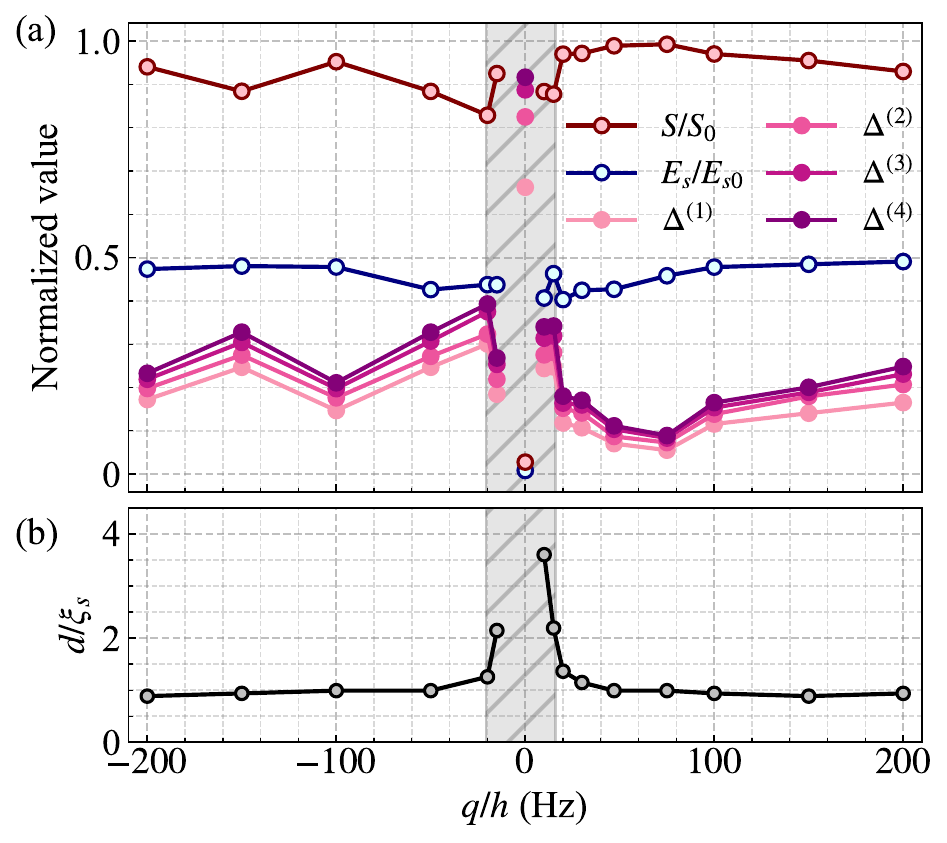}
    \centering
\caption{Effects of the quadratic Zeeman energy on the stationary turbulent state. (a) Normalized entropy $S/S_0$, spin interaction energy $E_s/E_{s0}$, trace distances $\Delta^{(k)}$ ($k=1,2,3$, $4$), and (b) domain size $d$ as functions of $q$. The quantities were evaluated at $t = 150t_s$. The gray-shaded region in both (a) and (b) indicates that the trace distance is not in a steady state at $t=150t_s$.  
}
    \label{FIG5}
\end{figure}

\begin{figure*}[t]
    \includegraphics[width=17.5cm]{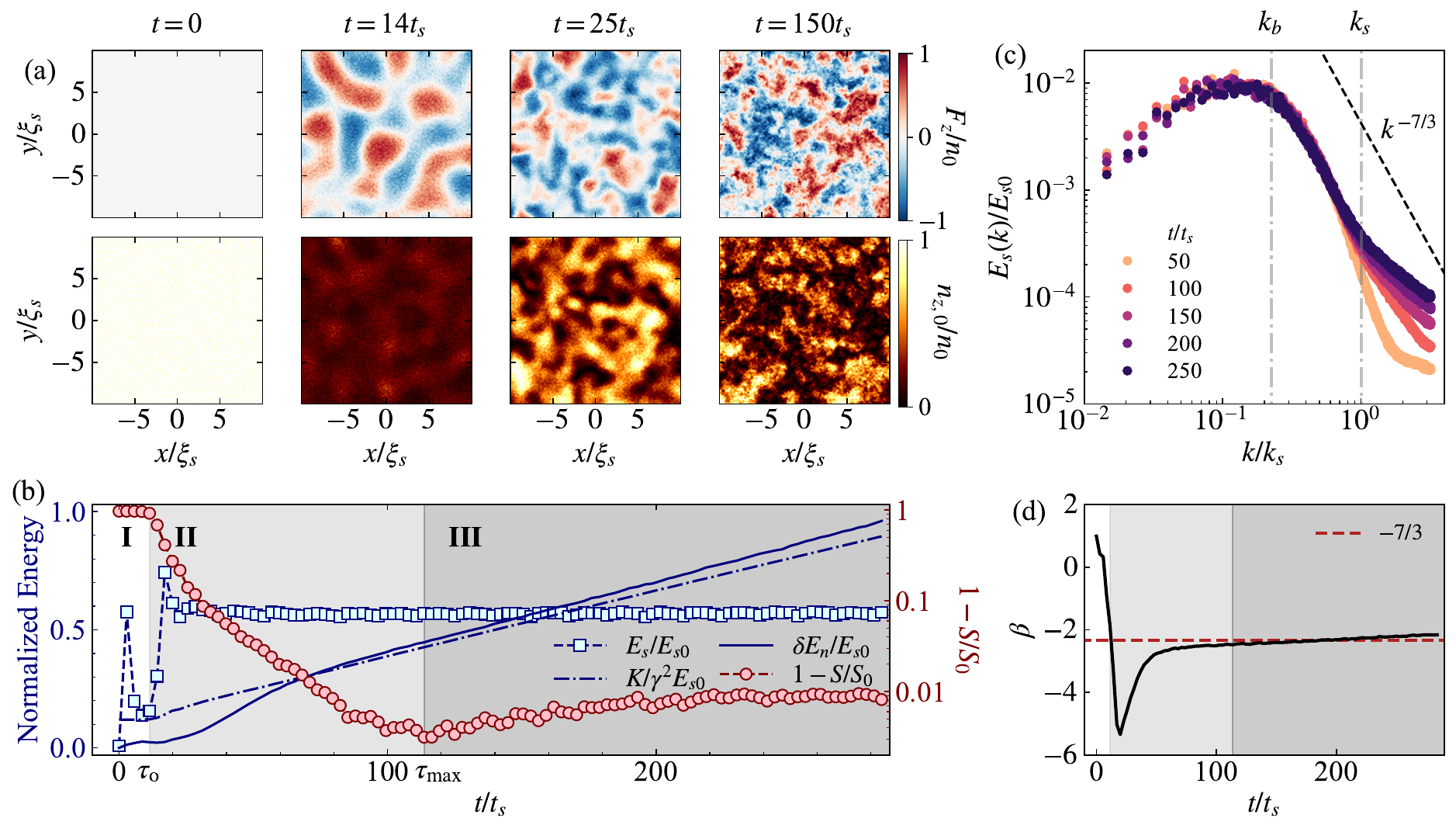}
    \centering
\caption{Turbulence in a spinor BEC with ferromagnetic spin interactions. (a) Distributions of magnetization $F_z(\boldsymbol{r})$ (upper) and $m_F$=0 spin component density $n_{z,0}(\boldsymbol{r})$ (lower) for different times. (b) Time evolution of spin interaction energy $E_s$ (blue square), kinetic energy $K$ (blue dashed), 
increment of density interaction energy $\delta E_n$ (blue solid), and spin entanglement entropy $S$ (red circle). The right $y$-axis uses a logarithmic scale for $1-S/S_0$. Three stages \textbf{I}, \textbf{II}, and \textbf{III} are indicated by the background colors as in Fig.~\hyperref[FIG1]{1(b)}. (c) Spin energy spectrum $E_s(k)$ at various times. The power exponent $\beta$, determined from a fit to the data in $k_b<k<k_s$, is displayed as a function of time in (d). The red dashed line indicates $\beta=-7/3$.}
    \label{FIG6}
\end{figure*}

\begin{figure}[t]
    \includegraphics[width=8.6cm]{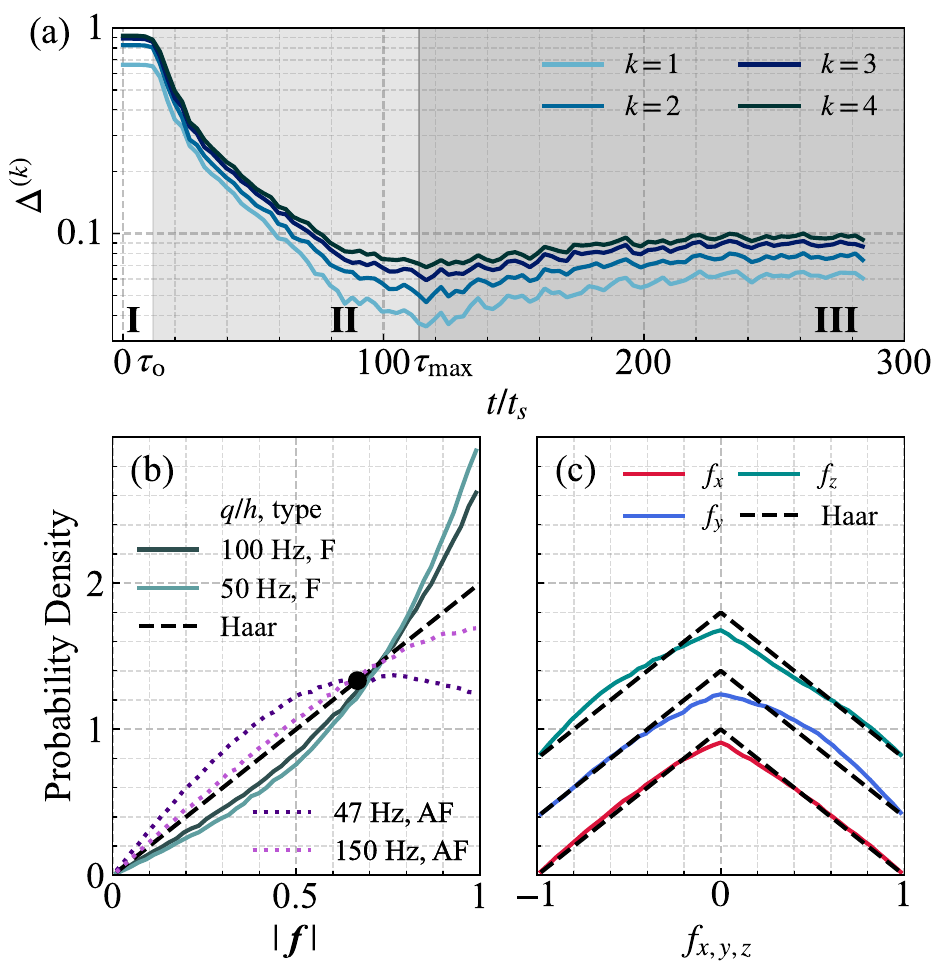}
    \centering
\caption{Spin statistics of the turbulent BEC with ferromagnetic spin interactions. (a) Trace distances $\Delta^{(k)}$ ($k=1,2,3,4$) between the $k$-th moments of $\mathcal{E}$ and the Haar random ensemble as functions of time. The $y$-axis has a logarithmic scale. (b) Probability density distributions of $\vert \boldsymbol{f} \vert$ and (c) $f_{x,y,z}$ at $t=150t_s$. The dashed lines indicate the corresponding distributions from the Haar random ensemble. In (b), `F' (`AF') on the caption corresponds to the ferromagnetic (antiferromagnetic) interactions. The black dot indicates $(2/3,4/3)$. In (c), the data for $f_y$ and $f_z$ were plotted with offsets of $0.4$ and $0.8$, respectively, for clarity.
}
    \label{FIG7}
\end{figure}

To visualize the spin randomization process, in Fig.~\hyperref[FIG4]{4(a)} we present the element distribution of the ensemble $\mathcal{E}$ in the coordinate space of $\boldsymbol{f}=(f_x,f_y,f_z)=\boldsymbol{F}/n$ at different times. Initially, the elements of $\mathcal{E}$ are localized near $\boldsymbol{f}$=0 and as time passes, they spread, eventually covering most of the whole parameter space with $\vert \boldsymbol{f} \vert \leq 1$. 

The probability density of the normalized spin vector magnitude $\vert \boldsymbol{f} \vert$, $P^{\mathcal{E}}_{|\boldsymbol{f}|}(\eta=\vert \boldsymbol{f} \vert)$, is evaluated for the ensemble $\mathcal{E}$ at $t=284 t_s$, as shown in Fig.~\hyperref[FIG4]{4(c)}. For comparison, we also plot the corresponding probability density distribution of the Haar ensemble, $P^{\text{Haar}}_{|\boldsymbol{f}|}(\eta)$=$2\eta$, whose derivation is provided in Appendix A. The ensemble $\mathcal{E}$ has more population on small $|\boldsymbol{f}|$. We attribute it to the antiferromagnetic interactions of the system, which energetically favor small $|\boldsymbol{f}|$. When the density fluctuations are not significant, the spin interaction energy is related to the second moment of the spin vector magnitude as $E_s \approx E_{s0} \langle \vert \boldsymbol{f} \vert^2 \rangle_\mathcal{E}$ from Eq.~(\ref{Energy}), with $\langle \vert \boldsymbol{f} \vert^2 \rangle_{\mathcal{E}} =\int d\eta P^{\mathcal{E}}_{|\boldsymbol{f}|}(\eta) {\eta}^2$.\footnote{According to Eq.~(\ref{observable}), $\langle \vert \boldsymbol{f} \vert^2 \rangle_{\mathcal{E}}=\text{f}_x^{(2)}+\text{f}_y^{(2)}+\text{f}_z^{(2)}$.} We obtain  $\langle \vert \boldsymbol{f} \vert^2 \rangle_{\mathcal{E}}=0.43$, which is consistent with the measured value of $E_s/E_{s0}$, whereas $\langle \vert \boldsymbol{f} \vert^2 \rangle_{\text{Haar}} = 0.5$ for the Haar ensemble. 
We may define the spin interaction energy for the Haar random ensemble as $E_s^{\text{Haar}}= 0.5 E_{s0}$.

Additionally, we examine the probability density profile of the spin vector component $f_\alpha$, $P^{\mathcal{E}}_{f_\alpha}(\eta=f_\alpha)$. Our numerical results are displayed in Fig.~\hyperref[FIG4]{4(d)}, together with the corresponding density profile of the Haar ensemble, $P^{\text{Haar}}_{f_\alpha}(\eta)$=$1$$-$$\vert \eta \vert$ (see Appendix A). In the $f_{x,y,z}$=0 region, $P^{\mathcal{E}}_{f_\alpha}$ is slightly higher than $P^{\text{Haar}}_{f_\alpha}$, which is consistent with the observed deviation of $P^{\mathcal{E}}_{|\boldsymbol{f}|}$ from $P^{\text{Haar}}_{|\boldsymbol{f}|}$ in Fig.~\hyperref[FIG4]{4(c)}. Note that the probability density profile $P^{\mathcal{E}}_{f_\alpha}(\eta)$ is directly accessible in experiments through magnetization imaging of the BEC along the spin axis $\alpha$~\cite{SeoHQV}.

%\subsection{Length scale of spin texture}

\subsection{Quadratic Zeeman effect}

In a spinor BEC system, the quadratic Zeeman energy $q$ introduces spin anisotropy and plays a critical role in determining the spin ground state, while also competing with the effects of spin interactions~\cite{kawaguchi2012spinor}. In the previous experiment in \cite{hong2023spin}, it was observed that the characteristic length scale of the spin texture increases as the quadratic Zeeman energy $q$ decreases. To further investigate the impact of $q$ on the randomness of the turbulent spin texture, we perform numerical simulations for different $q$ values, including the $q<0$ regime where the system's ground state is the easy-plane polar phase in the absence of magnetic driving.

In Fig.~\hyperref[FIG5]{5}, we present our numerical results for various $q$ values at $t=150t_s$, plotting the entropy $S$, spin interaction energy $E_s$, trace distances to quantum state $k$-design ($k=1,2,3,4$), and domain size $d$. Our observations reveal that spin randomization becomes more efficient when $q/h\approx 75$~Hz, at which point $S$ is maximized and $\Delta^{(k)}$ is minimized. However, we also note that $E_s$ increases with increasing $q$, approaching the value $E_s^{\text{Haar}}$ of the Haar ensemble. The size of the domain increases as $\vert q \vert$ decreases, in agreement with the experimental results. It is important to mention that within the range of $-20$~Hz$<q/h<$15~Hz (the shaded region), we observed that the system did not reach a steady state at $t=150t_s$, indicating a longer relaxation time for small $q$. Furthermore, for the specific case of $q=0$, no turbulence was generated, highlighting the critical role of $q$ in the generation of stationary turbulence. However, the underlying mechanisms responsible for the efficiency of magnetic driving in randomizing the spin texture remain unclear, which warrants further investigation.

\subsection{Ferromagnetic spin interactions}

Finally, we extend our study to a case with ferromagnetic interactions by changing the sign of $g_s$ to have $g_sn_0/h=-45$~Hz. Here, we also change the quadratic Zeeman energy to $q/h=100$~Hz because the initial EAP state with $\Psi_0 = \sqrt{n_0}(0,1,0)^\text{T}$ is dynamically unstable for $g_s<0$ and $q<2\vert g_s \vert n_0$, even without magnetic driving~\cite{sadler2006spontaneous,kawaguchi2012spinor}. 

In Fig.~\hyperref[FIG6]{6}, we present the numerical results for the ferromagnetic BEC system under magnetic driving. As in the previous case with antiferromagnetic interactions, turbulence with a complex spin texture is generated and sustained in the system for a long time. The time evolution of many characteristic quantities is similar to that observed in the antiferromagnetic case: $\tau_o=11 t_s$ and $\tau_\text{max}=114 t_s$, and $E_s$ converges to 0.57$E_{s0}$ [Fig.~\hyperref[FIG6]{6(b)}]. The spin energy is higher than $E_{s}^{\text{Haar}}$, which is due to the ferromagnetic interactions of the system. The spin energy spectrum $E_s(k)$ reveals the same power-law behavior as $\beta\approx -7/3$ [Figs.~\hyperref[FIG6]{6(c)} and \hyperref[FIG6]{6(d)}]. 

We also present the time evolution of the trace distances from the Haar ensemble in Fig.~\hyperref[FIG7]{7(a)}, demonstrating that the system in a stationary turbulent state is close to the quantum state 4-design. The probability density distribution of the magnitude of the density-normalized spin vector, $P^{\mathcal{E}}_{|\boldsymbol{f}|}(\eta)$, and that of the normalized magnetization, $P^{\mathcal{E}}_{f_\alpha}(\eta)$, are plotted in Figs.~\hyperref[FIG7]{7(b)} and \hyperref[FIG7]{7(c)}, respectively. Due to the ferromagnetic interactions, the probability for a high $|\boldsymbol{f}|$ is higher than that of the Haar ensemble. 

Intriguingly, we observe that at $|\boldsymbol{f}|\approx 0.7$, $P^{\mathcal{E}}_{|\boldsymbol{f}|}$ consistently shows similar values close to 1.4  across different signs of $g_S$ and various values of $q$. Moreover, we find that the profile can be well described by a quadratic function. To satisfy the conditions of $\int_0^1 d\eta P^{\text{fit}}_{|\boldsymbol{f}|}(\eta)=1$ and  $P^{\text{fit}}_{|\boldsymbol{f}|}(0)=0$, we propose the functional form as  
\begin{eqnarray}
    P^{\text{fit}}_{|\boldsymbol{f}|}(\eta) = b{\eta}(\eta-\frac{2}{3}) + 2\eta,
    \label{quadFit}
\end{eqnarray}
where a single parameter $b$ characterizes the probability density profile with a fixed point of $P^{\mathcal{E}}_{|\boldsymbol{f}|}(2/3)=4/3$. Furthermore, the relation $E_s \approx E_{s0} \langle \vert \boldsymbol{f} \vert^2 \rangle_\mathcal{E}$ suggests that $b$ can be approximated as $b\approx30 (E_s-E_{s}^{\text{Haar}})/E_{s0}$. Thus, $b<0$ for $g_s>0$ and $b>0$ for $g_s<0$, as observed, and the Haar random ensemble corresponds to $b=0$ with $E_s=E_s^{\text{Haar}}$.

\section{Summary and outlooks}

We conducted numerical investigations to characterize the spin texture in the stationary turbulent state of a driven spinor BEC. Our analysis revealed several key findings. First, through the analysis of entanglement entropy and magnetization correlation functions, we demonstrated the isotropic nature of the spin texture, highlighting its uniformity across different spatial directions. We also observed a $-7/3$ power-law behavior in the spectrum of the spin interaction energy, indicating the presence of turbulent dynamics in the system.

To further investigate the spin randomness of the spin texture, we derived a spin state ensemble using position projection. Comparing this ensemble to the Haar random ensemble, which serves as a reference for a fully random spin state, we found that the spin state ensemble closely approximates the quantum state 4-design. This suggests a high degree of spin randomness within the turbulent spin texture. Furthermore, we examined the probability density distribution of magnetization and discovered a peculiar functional form that can be parameterized by the system's spin interaction energy.

Our numerical study significantly improves our understanding of the characteristics of stationary spin turbulence in the spinor BEC system and provides support for previous experimental findings. However, it is important to acknowledge that the underlying mechanisms responsible for sustaining turbulent states are not yet clearly understood. In particular, it has been observed that in the absence of field fluctuations, the system relaxes to the ground state, as discussed in \cite{hong2023spin}. Exploring an expanded parameter space of magnetic driving, including driving strength $\Omega$, field fluctuation magnitude $\delta_0$, and frequency $\omega_\delta$, would be instrumental in unraveling the mechanisms that sustain the turbulent state under magnetic driving.

Finally, as a possible extension of this work, we consider a spinor condensate trapped in optical lattices. In this scenario, the notion of the projected spin ensemble becomes more relevant due to the presence of lattice sites. One intriguing possibility is to start from a Mott insulating phase where fluctuations of atom numbers for each lattice site are strongly suppressed. To mimic this situation, we performed preliminary studies by neglecting the kinetic energy in our numerical simulations. Surprisingly, we observed that a random spin ensemble can still be obtained through magnetic field driving. This observation suggests that the chaotic nature of the periodically driven spin-1 system plays a crucial role in the generation of the random spin ensemble~\cite{liu2022classical,cheng2010chaotic}, providing an interesting prospect for future experimental investigations.

\begin{acknowledgments}
This work was supported by the National Research Foundation of Korea (Grants No. NRF-2018R1A2B3003373 and No. NRF-2023R1A2C3006565) and the Institute for Basic Science in Korea (Grant No. IBS-R009-D1).
\end{acknowledgments}

\appendix

\section{Haar random ensemble and magnetization distributions for a spin-1 system}

According to Ippoliti and Ho~\cite{ippoliti2022dynamical}, the distribution of the projected ensemble is called \textit{Haar random ensemble} (i.e., uniformly or unitarily-invariant ensemble) if the statistics of the system has a maximally entropic distribution not just at the level of expectation values of local observables, but also at the level of the Hilbert space. Owing to the Schur-Weyl duality, the $k$-th moment of the Haar ensemble is given by
\begin{eqnarray}
    \rho_{\text{Haar}}^{(k)}&=& \int_{\phi\sim\text{Haar}(H_{\mathbb{S}})} d\phi~ (\phi \phi^{\dagger})^{\otimes k} \\
    &=& 2\frac{\sum_{\boldsymbol{\pi} \in \mathcal{S}_k}\text{Perm}(\boldsymbol{\pi})}{(k+2)!},
    \label{Haar}
\end{eqnarray}
where $\mathcal{S}_k$ is the symmetric group on $k$ elements. $\text{Perm}(\boldsymbol{\pi})$ is a representation of $\boldsymbol{\pi} \in \mathcal{S}_k$ on $k$ replicas of the Hilbert space $\mathcal{H}_{\mathbb{S}}$, which permutes the tensor products as $\text{Perm}(\boldsymbol{\pi}) \vert \phi_{1} \rangle \otimes \cdots \otimes \vert \phi_{k} \rangle = \vert \phi_{\boldsymbol{\pi}^{-1}(1)} \rangle \otimes \cdots \otimes \vert \phi_{\boldsymbol{\pi}^{-1}(k)} \rangle$~\cite{PRXQuantum.4.010311,harrow2013church}.

Using Eq.~(\ref{Haar}), $f_z^{(k)}$ for the Haar ensemble is expressed as follows. The spin states are denoted by $\vert + \rangle = (1,0,0)^{T}$, $\vert 0 \rangle = (0,1,0)^{T}$, and $\vert - \rangle = (0,0,1)^{T}$, giving $\text{f}_z = \vert + \rangle \langle + \vert - \vert - \rangle \langle - \vert$.
\begin{eqnarray}
    &f_{z,\text{Haar}}^{(k)}&=\text{tr}\big( \rho^{(k)}_\text{Harr} \text{f}_z^{\otimes k} \big)  \nonumber \\
    &=& \frac{2}{(k+2)!} \text{tr}(\sum_{\boldsymbol{\pi} \in \mathcal{S}_k} \text{Perm}(\boldsymbol{\pi}) (\vert + \rangle \langle + \vert - \vert - \rangle \langle - \vert)^{\otimes k}) \nonumber \\
    &=& \frac{2}{(k+2)!} \sum_{j=0}^{k} \frac{(-1)^{j}}{(k-j)!j!} \sum_{\boldsymbol{\pi},\boldsymbol{\pi'} \in \mathcal{S}_k} \text{tr} \Big[ \text{Perm}(\boldsymbol{\pi} \circ \boldsymbol{\pi'}) \nonumber \\
    &&(\vert + \rangle \langle + \vert)^{\otimes (k-j)} \otimes (\vert - \rangle \langle - \vert)^{\otimes j} \text{Perm}(\boldsymbol{\pi}'^{-1}) \Big] \nonumber \\
    &=&\frac{2}{(k+2)!} \sum_{j=0}^{k} \frac{(-1)^{j}}{(k-j)!j!} \sum_{\boldsymbol{\pi},\boldsymbol{\pi'} \in \mathcal{S}_k} \text{tr} \Big[ \nonumber \\
    && \text{Perm}(\boldsymbol{\pi}'^{-1} \circ \boldsymbol{\pi} \circ \boldsymbol{\pi'}) (\vert + \rangle \langle + \vert)^{\otimes (k-j)} \otimes (\vert - \rangle \langle - \vert)^{\otimes j} \Big] \nonumber \\
    &=&\frac{2}{(k+2)!} \sum_{j=0}^{k} \frac{(-1)^{j}}{(k-j)!j!} \sum_{\boldsymbol{\pi''},\boldsymbol{\pi'} \in \mathcal{S}_k} \text{tr} \Big[ \text{Perm}(\boldsymbol{\pi''}) \nonumber \\
    && (\vert + \rangle \langle + \vert)^{\otimes (k-j)} \otimes (\vert - \rangle \langle - \vert)^{\otimes j} \Big] \nonumber \\
    &=&2\frac{\text{ord}(\mathcal{S}_k)}{(k+2)!} \sum_{j=0}^{k} \frac{(-1)^{j}}{(k-j)!j!} \sum_{\boldsymbol{\pi''} \in \mathcal{S}_k} \text{tr} \Big[ \text{Perm}(\boldsymbol{\pi}'') \nonumber \\
    && (\vert + \rangle \langle + \vert)^{\otimes (k-j)} \otimes (\vert - \rangle \langle - \vert)^{\otimes j} \Big] \nonumber \\
    &=& \frac{1-(-1)^{k+1}}{(k+2)(k+1)}.
    \label{magDistributionDerive}
\end{eqnarray}
From the relation
\begin{equation}
    f_z^{(k)}=\int_{-1}^{1} P^{\mathcal{E}}_{f_z} (\eta) {\eta}^k d\eta, 
\end{equation}
where $P_{f_z}^{\mathcal{E}}(\eta)$ is the probability distribution of $f_z$ for the ensemble $\mathcal{E}$, the result of Eq.~(\ref{magDistributionDerive}) provides the bilateral Laplace transform of the probability distribution as 
$\mathcal{B}\{P^{\text{Haar}}_{f_z}\} (s) = 2(\cosh{s}-1)/s^2$, where $\mathcal{B}\{{h}\}(s) = \int_{-\infty}^{\infty} h(\eta)e^{-\eta s} d\eta$, yielding 
\begin{eqnarray}
    P^{\text{Haar}}_{f_z}(\eta) = (1-\vert \eta \vert)~\Theta(1-\vert \eta \vert),
    \label{magDistribution}
\end{eqnarray}
where $\Theta$ is the step function. Given that Haar ensemble is isotropic, the expression is generalized to arbitrary $\alpha$ axis. Using the relation of Eq.~(\ref{Relation}), we obtain the probability distribution of the spin vector strength $\vert \boldsymbol{f} \vert$ for the Haar ensemble as 
\begin{eqnarray}
    P^{\text{Haar}}_{|\boldsymbol{f}|}(\eta) =2\eta & (0\leq \eta \leq 1).
    \label{F_distribution}
\end{eqnarray}

\section{Relation of probability densities}

When the ensemble is isotropic in the spin direction, i.e., $f_\alpha^{(k)}= f_z^{(k)}$ for all $\alpha$ and $k$, the following relation holds,
\begin{eqnarray}
    |f_z|^{(k)}&=&\int_{0}^{1} 2P^{\mathcal{E}}_{f_z} (\eta) {\eta}^k d\eta \\
&=& \frac{1}{4\pi}\int_{\eta=0}^{1}\int_{\theta=0}^{\pi} P^{\mathcal{E}}_{|\boldsymbol{f}|} (\eta) \vert \eta\cos{\theta} \vert^{k} 2\pi \sin{\theta} d\theta d\eta \nonumber \\
 &=& \frac{1}{k+1} \int_{\eta=0}^{1} P^{\mathcal{E}}_{|\boldsymbol{f}|} (\eta) {\eta}^{k} d\eta \nonumber
\\
 &=& \frac{1}{k+1} |\boldsymbol{f}|^{(k)}
   \label{momentRelation}
\end{eqnarray}
Then, the unilateral Laplace transforms, related to the moment-generating function, of the two probability density functions can be expressed as follows:
\begin{eqnarray}
    2\mathcal{L}\{P^{\mathcal{E}}_{f_z}\} (s) + 2s\frac{d}{ds} \mathcal{L}\{P^{\mathcal{E}}_{f_z}\} (s) = \mathcal{L}\{ P^{\mathcal{E}}_{|\boldsymbol{f}|} \} (s),
\end{eqnarray}
where $\mathcal{L}\{{h}\}(s) = \int_{0}^{\infty} h(\eta)e^{-\eta s} d\eta$.
Taking the inverse Laplace transformation, we obtain
\begin{eqnarray}
    2P^{\mathcal{E}}_{f_z}(\eta) - 2\frac{d}{d\eta}[\eta P^{\mathcal{E}}_{f_z}(\eta)] = P^{\mathcal{E}}_{|\boldsymbol{f}|}(\eta), \nonumber \\
    \frac{d}{d\eta}P^{\mathcal{E}}_{f_z}(\eta) = -\frac{1}{2\eta} P^{\mathcal{E}}_{|\boldsymbol{f}|}(\eta), 
    \label{Relation}
\end{eqnarray}
for $\eta \geq 0$. 
%$p_m(r) = p_m(\vert r \vert)$ for $-\infty < r < \infty$.
This is the general relation where the ensemble is isotropic.

For the probability function $P^{\text{fit}}_{|\boldsymbol{f}|}(\eta)$ in Eq.~(\ref{quadFit}), Eq.~(\ref{Relation}) yields 
\begin{eqnarray}
    \frac{d}{d\eta} P^{\text{fit}}_{f_\alpha}(\eta) = -\frac{1}{2}b\eta-(1-\frac{1}{3}b), & 0 \leq \eta \leq 1.
\end{eqnarray}
Given that $\int_{-1}^{1} P^{\text{fit}}_{f_\alpha}(\eta) d\eta =1$ and $P^{\text{fit}}_{f_\alpha}(-\eta) = P^{\text{fit}}_{f_\alpha}(\eta)$ for $\vert \eta \vert \leq 1$, we obtain
\begin{eqnarray}
    P^{\text{fit}}_{f_\alpha}(\eta) = -\frac{b}{4} (1-\vert \eta \vert)^2 + (1+\frac{b}{6}) (1- \vert \eta \vert). 
    \label{magQuad}
\end{eqnarray}

%\bibliography{ref}
%

\end{document}